# Electronic band structure of α-(Per)$_2$M(mnt)$_2$ compounds

Enric Canadell [1], Manuel Almeida [2], James Brooks [3]

[1] Institut de Ciència de Materials de Barcelona, (CSIC), Campus Universitari de Bellaterra, E-08193 Bellaterra, Spain

[2] Instituto Tecnológico e Nuclear / CFMCUL, Estrada Nacional nº 10, P-2686-953 Sacavém, Portugal

[3] NHMFL/Physics, Florida State University, Tallahassee Florida, 32310, USA

**Abstract**

The band structure of α-(Per)$_2$M(mnt)$_2$, leads to a description of these materials as nearly perfectly one dimensional conductors. The conduction is mainly along the stacking direction of the partially oxidized perylene molecules, (Per)$^{½+}$, with virtually no interchain bandwidth. However, recent high magnetic field experiments suggest orbital coupling of the magnetic field to the electronic structure, indicating a finite interchain bandwidth. The details of the band structure and the possible variances, at low energies, from a perfect one dimensional system are examined. In particular, multiple quasi-one dimensional Fermi surface sheets, which would become important at low temperatures, may lead to an explanation for the experimental finding of a magnetic field induced high field charge density wave ground state. Experimental tests to observe the effects of the finite interchain bandwidth are proposed.

The α-phases of the (Per)$_2$M(mnt)$_2$ compounds, where Per = perylene, mnt = maleonitriledithiolate and M = Au, Pt, Pd, Ni, Cu, Co, Fe have attracted continuous interest for more than 20 years, since the report on the Pt compound [1], where a quasi-one-dimensional (1D) metallic behaviour was found for the first time in this family of molecular solids. Subsequent work demonstrated that these compounds belong to an almost isostructural family. Here conducting stacks of partially oxidized perylene molecules coexist with anionic stacks, which for some transition metal complexes (e.g., Ni, Pt, Pd and Fe) may have localized magnetic moments [2,3,4]. These two types of chains, are both prone to the instabilities typical of conducting and magnetic 1D systems as demonstrated by diffuse x-ray studies [5,6] and the possible coupling of these instabilities has been at the heart of different studies on these materials [2,3,7].

The conducting chains of perylene molecules undergo at low temperatures (8, 12, 25, 28, 32, 58, 73 K for Pt, Au, Ni, Pd, Cu, Co and Fe, respectively) a Peierls transition to a charge density wave (CDW) ground state, where clear non-linear electrical transport properties are observed [8,9,10,11,12]. For M = Ni, Pt and Pd, the magnetic chains undergo simultaneously an additional spin-Peierls transition associated with a dimerisation of the M(mnt)$_2$ magnetic chains [2,3,4,7,13]. It has been proposed that the antiferromagnetic coupling between the M(mnt)$_2^-$ paramagnetic units is mediated by the conduction electrons in the perylene chains through a RKKY type mechanism [14,15].

The room temperature crystal structure of the Au, Pt, Pd, Ni, Cu compounds belongs to the space group P2$_1$/n with Z = 2 and it contains regular chains along **b** of both donor and acceptor units [1,16,17,18]. The asymmetric unit contains a full perylene molecule and half an anion with the metal located at a center of symmetry while each unit cell contains 4 perylene units and two anions which are related by a screw axes. The perylene sublattice, which will be the focus of the present work, is shown in fig. 1. The M(mnt)$_2$ stacks are located on the open channels of Figure 1. The perylene stacks alternate in the **a,c** plane with chains of M(mnt)$_2$ anions in such a way that each stack of anions is surrounded by six stacks of perylene and each perylene stack has 3 stacks of perylene and 3 stacks of anions as nearest neighbours.

These compounds show a room temperature conductivity of the order of 700 S/cm along the high conductivity **b** axis and the anisotropy in the **a,b** plane is estimated to be 900. The bandwidth estimated from both thermopower and magnetic susceptibility measurements, is in the range 0.4-0.6 eV [4,17,18,19] depending on the metal M. These values are in reasonable agreement with theoretical calculations based on the extended Huckel approach, which indicates an intrachain transfer integral ($t_b$) of 0.147 eV [20]. This type of structure, consisting in a tight packing of planar aromatic donors, with the graphite like overlapping mode depicted in fig. 2, is expected to lead to very anisotropic properties including an almost purely 1D behaviour with negligible interstack interactions of the order of 1 meV. In fact the presence of hydrogen atoms in the periphery of the perylene molecules, which give no contribution to the highest occupied molecular orbital (HOMO; see fig. 2), render the electronic interactions mediated by the interchain hydrogen contacts ineffective. Hence the system was for long time expected to behave as almost purely 1D metal, and the Fermi surface was predicted to be, in good approximation, two parallel, almost flat sheets with Fermi momentum in the **b**-axis direction.

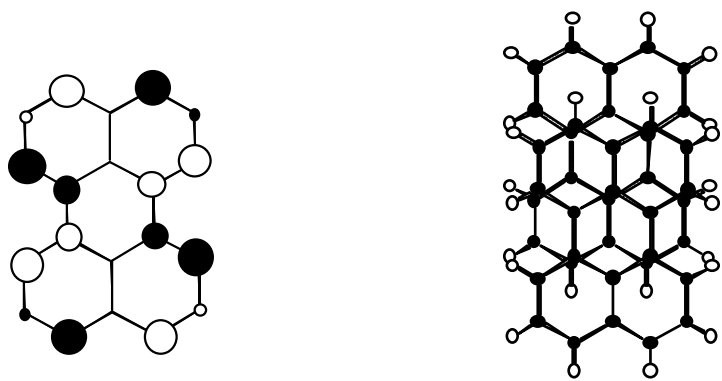

Figure 2. Left: Schematic representation of the perylene HOMO. Right: overlap mode of the perylene donors along the **b**-axis.

More recently, studies of the Au and Pt compounds under large magnetic fields allowed the full suppression of the CDW state [21]. In the Pt compound where the transition temperature is lower, this suppression is followed by a cascade of field induced transitions at higher fields [22] .Both the suppression of the CDW state and the appearance of the field induced transitions is dependent on the magnetic field orientation, showing that the magnetic field effect involves an orbital mechanism. In spite of the resemblance between the field induced transition in (Per)$_2$M(mnt)$_2$ compounds with the FISDW transitions previously described in the Bechgaard salts, the transitions in the perylene compounds cannot be fully described by the standard models for the CDW under field [22,23]. At low temperature, where both the CDW and field induced transitions are observed, details of the Fermi surface below 1 meV are expected to become significant. Thus, these recent findings lead us to reevaluate the band structure of these compounds and to estimate the magnitude of the interchain interactions in order to have a more detailed description of the Fermi surface topology.

The calculations used an extended Huckel Hamiltonian [24] using both single-$\zeta$ and double-$\zeta$ type atomic orbital basis sets. This kind of approach has been very successful in predicting and rationalizing the electronic structure of molecular conductors [25]. In our calculations we used the room temperature crystal structure of $(Per)_2Pt(mnt)_2$ where, as sketched in fig. 1, there are four different types of interchain Per$\cdots$Per interactions between the three distinct chain pairs. Because of the screw axes, which relate two pairs of perylene stacks, the two interactions noted $t_1$, as well as the two interactions noted $t_2$, are identical. This is not the case for the interactions noted $t_3$ and $t_4$, between stacks not related by a screw axis. The transfer integrals calculated using both the single-$\zeta$ and double-$\zeta$ basis sets are listed in table 1. All the interchain interactions are indeed very small compared with the intrastack interaction (t) and the material should be very 1D.

Table 1. Calculated single-$\zeta$ transfer integrals for $(Per)_2Pt(mnt)_2$. Values in parenthesis correspond to double-$\zeta$ values.

| Interaction | Transfer integral (meV) |
|---|---|
| t | 148.6 (353.8) |
| $t_1$ | 1.7    (8.3) |
| $t_2$ | 2.4    (7.6) |
| $t_3$ | -0.2    (-1.1) |
| $t_4$ | 0.0    (0.0) |

The calculated single-$\zeta$ values should provide the more realistic values. The double-$\zeta$ results are probably an overestimation. This kind of calculation is usually carried out for systems more 2D, like many BEDT-TTF salts, in which many of the contacts are lateral and thus, where the overlap of the $p_z$ orbitals is less efficient [26]. The atomic orbitals are represented as a combination of two orbitals, one considerably more diffuse than the other, so that the overlaps are increased with respect to those of the calculations with a single-$\zeta$ basis set. In general, the real values of the transfer integrals will be in between those of the two types of calculations, 1D systems being more on the side of the single-ζ calculations and 2D systems more on the side of the double-ζ calculations. Taking into account the good agreement between the bandwidth estimates and 4t for the single-ζ calculations we tend to believe that the single-ζ values provide the better estimation. However, the double-ζ calculations suggest that even if some uncertainty remains in the absolute values of the transfer integrals, the ratio between the strengths of the interactions along different directions should remain approximately correct. In spite of the low values for interchain interactions they will become significant at low temperatures. Namely for the Pt compound, with a lower critical temperature of $T_c$ = 8 K, details smaller than 1 meV are expected to become relevant at temperatures approaching $T_c$. In this respect it should be kept in mind that the full details of the field induced transitions, which are expected to be sensitive to Fermi surface topology only become visible below 2K. In particular, it is clear from the values of the transfer integrals that the interchain interactions, in spite of being much smaller than the intrachain values, present a significant anisotropy in the **a,c** plane, since they are essentially negligible along the **c**-direction.

Some details of the band structure of $(Per)_2M(mnt)_2$ are shown in fig. 3. Since there are four perylene stacks per unit cell, there must be four HOMO bands crossing the Fermi level. Without the transverse interactions $t_1$, $t_2$, $t_3$ and $t_4$, the Fermi surface would be the superposition of four planes at $k_{b*}$ = ±0.375, but because of these transverse interactions, the Fermi surface splits into four pairs of sheets with some warping. According to our results this warping should be maximum around the **a\***-direction and minimum around the **c\***-direction. The nature of the band structure at around the Fermi level is shown enlarged (and not to scale) in fig. 3b, where the four bands can be clearly seen (for the meaning of the labels in the band structure see fig. 3c and the legend). In the Γ to Y direction it seems that there are just two bands because every band is really two superposed (practically degenerate) bands. This is due to the absence of interaction along **c** that makes the two pairs of bands practically identical. However the two bands are distinguished when directions perpendicular to the **b\***-direction are considered (Figure 3b). The gap between the two pairs of bands at Γ is due to the interchain interactions (thus, those along **a**). As can be seen, the effect of the small interchain interactions decreases when $k_{b*}$ increases. Because of the inclination of the molecules with respect to **b**, the band dispersion in the sections perpendicular to **b*** are not completely independent of the $k_{b*}$ component.

An estimate of the warping for the Fermi surface is given by the calculation of the band structure for a fixed value of the $k_{b*}$ component (0.375), which should approximately correspond to the Fermi wave vector along the chains, and different values of the $k_{a*}$ and $k_{c*}$ components (see fig. 3b). It may look surprising that in Figure 3b there is a gap separating the two upper and two lower bands. However note that because of non nil interchain interactions, the $k_{b*}$ component at the Fermi level is 0.375-δ for one pair of bands and 0.375+δ for the other pair (δ being very small, of the order of ~0.001-0.002), whereas the dispersion in fig. 3b is calculated for exactly 0.375. In the enlarged diagram, the dispersion along lines parallel to the **c***-direction (i.e., B → C and A → D) is practically nil confirming the absence of interactions along *c*. In contrast, there is some dispersion along the directions parallel to the **a***-direction ((i.e., A → B and D → C) because there are interchain interactions along **a**.

From these calculations it seems that the warping of the sheets can be of the order of 2 meV. Hence the Fermi surface contains four sheets with this kind of warping along **a*** and practically zero along **c*** as schematically depicted in fig. 4. With four sheets and relatively weak interchain interactions, the four Fermi surface sheets may interpenetrate each other, and by hybridization between the different sheets there could be regions with closed pockets. Although these effects usually would be ignored, at very low temperatures they may become important, as the experimental results suggest.

Because of the good agreement of the calculated dispersion along **b** and experimental estimations it would be quite surprising if the interchain interactions were grossly erroneous. The details of the topology of the band structure (even the very small changes) can be understood quite cleanly in terms of the number and relative inclination of the molecules and the values of the transfer integrals so that the same topology of the band structure (and Fermi surface) would be obtained if the interchain interactions were a little bit stronger. The **a,c** anisotropy predicted by these calculations is in gross qualitative agreement with experimental findings of magnetoresitence anisotropy [21,22]. However the actual values for the transfer integrals depend on several factors which usually are ignored. When the interactions are so small, the quality of the crystal structure, the effect of the thermal contraction (which should at least slightly increase the interactions), and the role of the anions may become important and somewhat change the coupling. Therefore low temperature X-ray studies will be needed in order to accurately compute alterations in the band structure that occur due to thermal contraction. Although the present calculations were carried out for the Pt compound, the differences with the Au compound should be very small.

Finally we should mention that even after taking into account possible low temperature structural modifications to be evaluated from low temperature X-ray diffraction experiments, the details of the band structure calculation and the magnitude of the Fermi surface warping ultimately must be tested experimentally. We conclude by considering possible experimental tests which can probe the details of the Fermi surface and the magnitude of the interchain bandwidth. For quasi-one dimensional organic conductors, angular dependent magnetoresistance effects may be used to test for details of the electronic structure that are due to interchain coupling. Most common is the so-called Lebed-type effect [27,28] where the magnetic field is rotated in the plane parallel to the **a,c** plane. Here, for integer ratios of the **a** and **c** lattice parameters, the magnetoresistance will exhibit peaks or dips. A similar method is to use the so-called Danner-Chaikin method [29], which allows the measurement of the interchain bandwidth. In this case the field is directed at an angle to the chain axis. It is necessary to be in the metallic state of the material, so in this case high fields and/or high temperatures must be used in order to avoid the CDW ground state and also the field induced CDW state. Another method is the microwave resonance probe (so-called periodic orbit resonance [30]) which can measure trajectories on the Fermi surface sheets. Also, if small pockets are present, field modulation measurements to measure the Shubnikov-de Haas effect for very small quantum oscillation frequencies can be effective [31]. These experiments are expected to confirm these Fermi surface calculations in the future.

This work is supported by NSF 02-03532; the NHMFL is supported by the National Science Foundation and the State of Florida. Work in Portugal is supported by FCT under contract POCT/FAT/39115/2001. Work at ICMAB is supported by DGI-Spain (Project BFM2003-03372-C03) and Generalitat de Catalunya (Project 2001 SGR 333).

**Figure captions**

Figure 1. View along the stacking axis **b** of the perylene sublattice in (Per)$_2$M(mnt)$_2$ and side views of different pairs of next neighbouring perylene stacks, with labeling of the different interchain transfer integrals.

Figure 2. Left: Schematic representation of the perylene HOMO. Right: overlap mode of the perylene donors along the **b**-axis

Figure 3. Calculated band structure of (Per)$_2$Pt(mnt)$_2$. a) Band dispersion along $\mathbf{k_{b*}}$. b) Detail of the four bands around the Fermi level, with energy dispersion at $\mathbf{k_{b*}}$ = 0.375 and different values of the $\mathbf{k_{a*}}$ and $\mathbf{k_{c*}}$ indicated in c). c) Section with $\mathbf{k_{b*}}$ = 0.375 with labeling of special reciprocal space points. Γ = (0, 0, 0), Y (0, 1/2, 0), A = (0, 0.375, 0), B = (1/2, 0.375, 0), C = (1/2, 0.375, 1/2), D = (0, 0.375, 1/2) and E = (-1/2, 0.375, 1/2) in units of the monoclinic reciprocal space vectors.

Figure 4. Estimated Fermi surface of (Per)$_2$Pt(mnt)$_2$. Note that this is a largely enlarged picture around $\mathbf{k_{b*}}$ = 0.375.

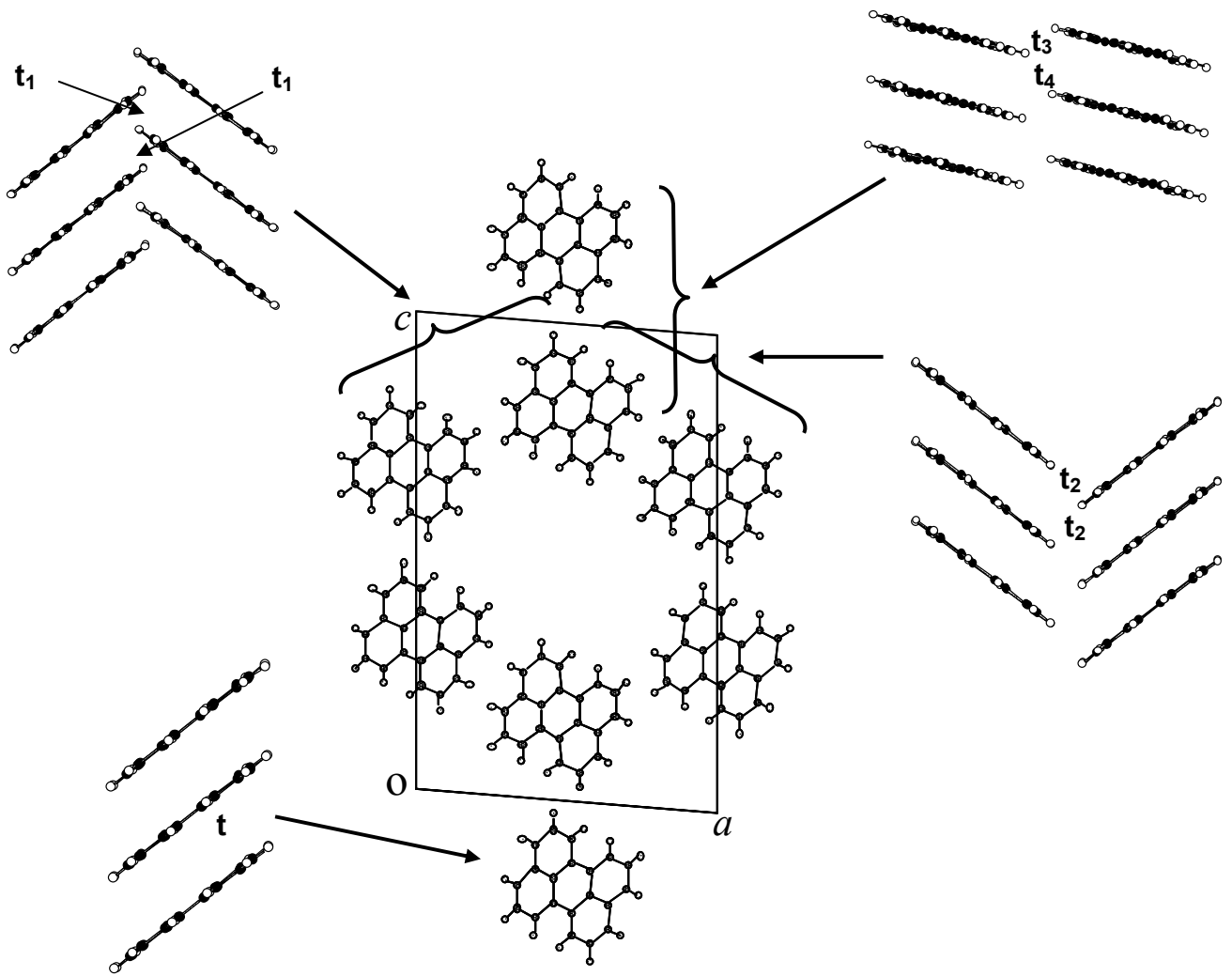

**Figure 1**

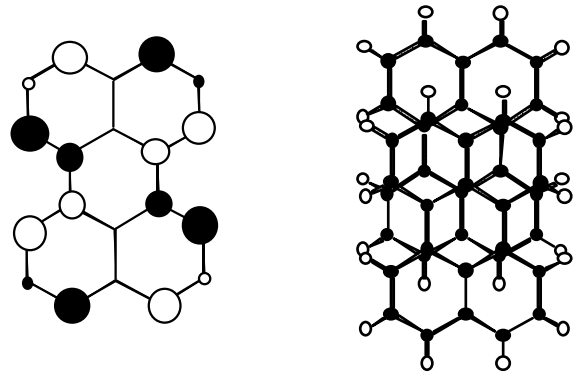

Figure 2. Left: Schematic representation of the perylene HOMO. Right: overlap mode of the perylene donors along the **b**-axis.

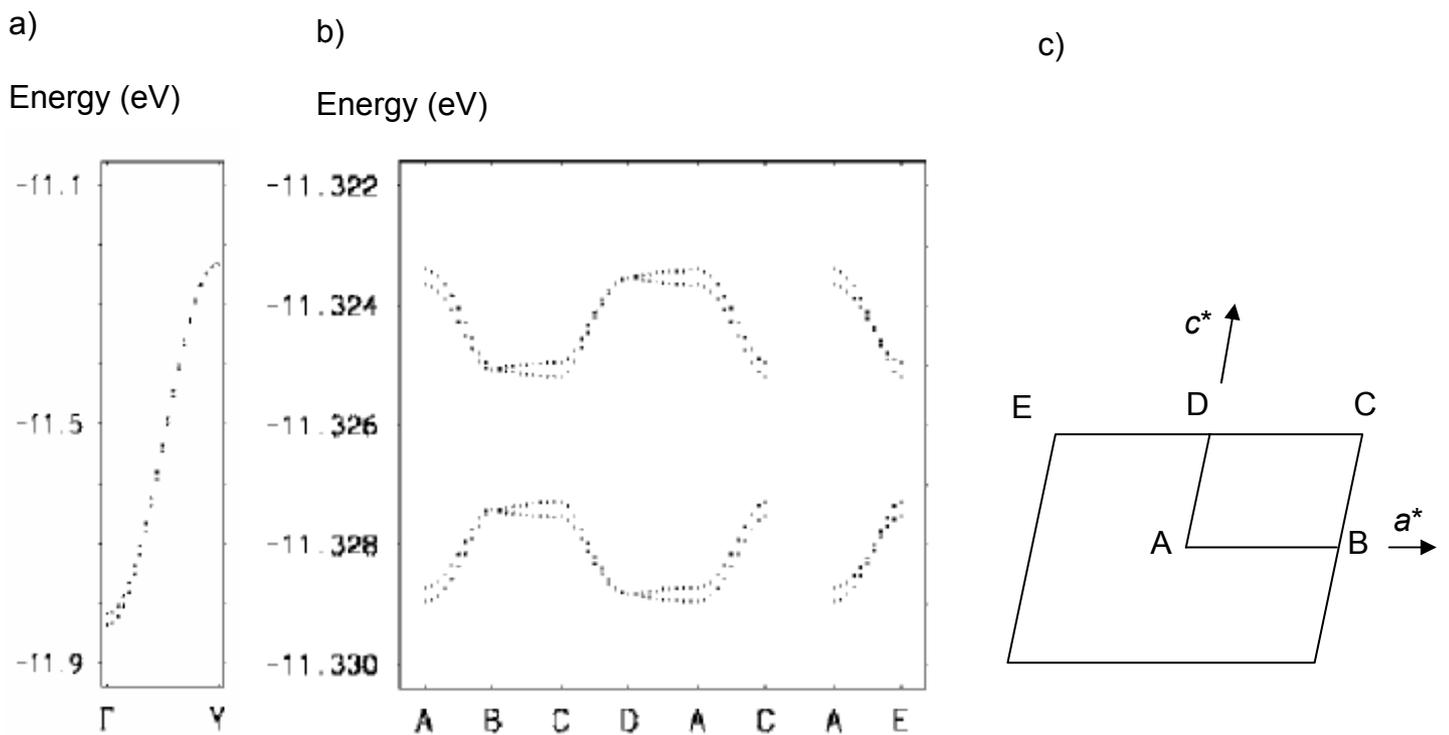

**Figure 3**

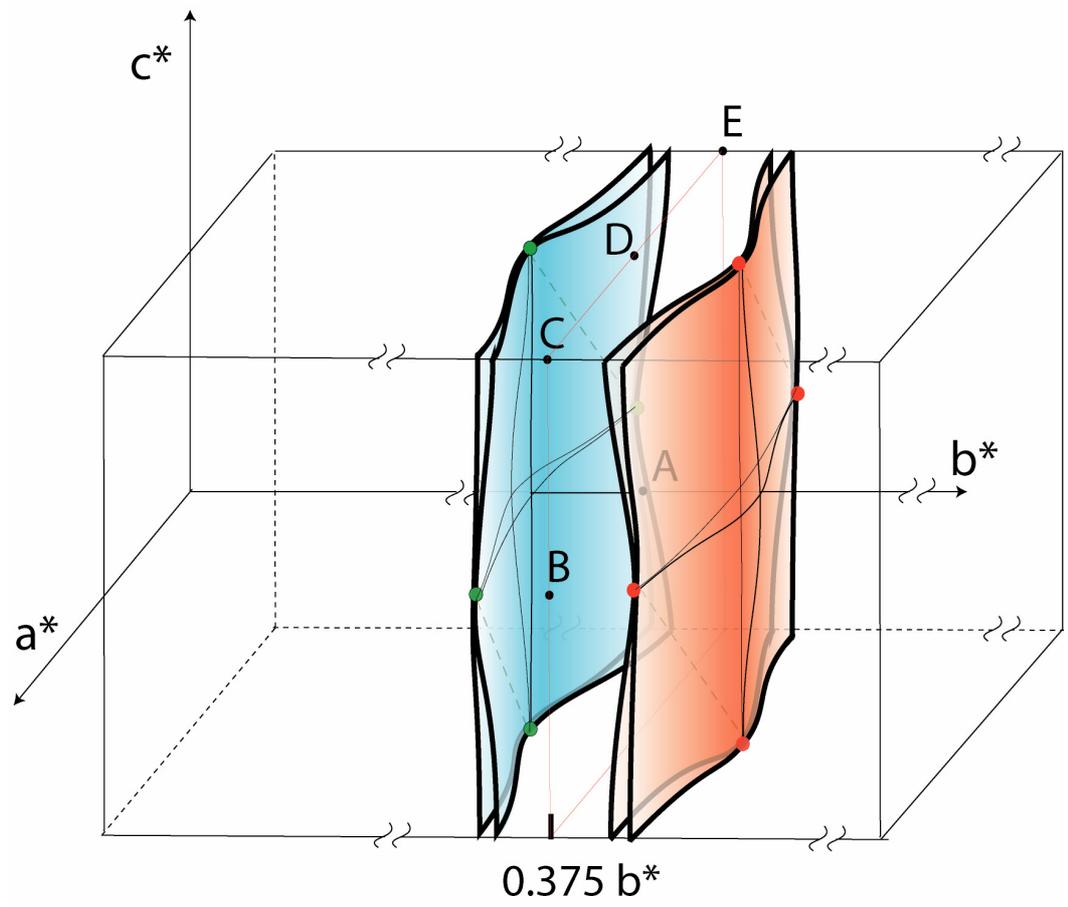

**Figure 4**